\documentclass[11pt,twoside]{article}

\usepackage{asp2006}
\usepackage{epsf}
\usepackage{psfig}
\usepackage{lscape}
\usepackage{natbib}
\usepackage{graphicx}

\begin{document}
\title{X-ray Observables of Magnetically Dominated Cavities in Clusters}   %%% Fill in title
\author{Steven Diehl, Masanori Nakamura, Hui Li}   %%% Fill in author names
\affil{Los Alamos National Laboratory, P.O. Box 1663, Los Alamos, NM, 87545}    %%% Fill in author affiliations
\author{and David Rafferty}   %%% Fill in author names
\affil{Ohio University, Clippinger Laboratories 251B, Athens, OH, 45701}    %%% Fill in author affiliations

\begin{abstract} %%% Abstract to run on from here.
%Don't forget the abstract. Also a reminder: 5 pages max for posters. Also remember to send stuff to Brian McNamara, as he wants to include the results into his review. 
We present results from magneto-hydrodynamic jet simulations in the magnetically dominated regime and find that they inflate cavities into the intracluster medium. Mock {\it Chandra} X-ray observations of our simulations show a strong resemblance to real cavities observed in X-ray images of clusters, both in  terms of their morphology and thermodynamic structure. An analysis of the evolution of bubble sizes in the multi-cavity system Hydra A, as well as in a large sample of 64 cavities in 32 clusters shows that bubbles tend to expand much faster than expected in the purely adiabatic regime. Instead, we find that the bubbles follow more closely a trend predicted by our current-carrying jet models.
\end{abstract}

%%% MAIN BODY OF TEXT GOES HERE. CONSULT "INSTRUCTIONS FOR AUTHORS USING
%%% LATEX2E MARKUP", SECTIONS 2.3-2.6 FOR HELP WITH EQUATIONS, FIGURES,
%%% AND TABLES.

\section{Introduction}   %%% Top level section head (remove "%" symbol)

The cooling time of the hot gas in the center of clusters is significantly shorter than the age of the cluster. Thus, the gas should have already cooled completely, dropped out of the X-ray phase and formed stars. This so-called Òcooling flow problemÓ suggests that something is heating the gas, and effectively preventing it from cooling. 
Chandra observations of clusters reveal common features in the X-ray emission that point toward the central AGN to have a significant impact. Images often show deficits in the X-ray emission - generally referred to as ÒcavitiesÓ or ÒbubblesÓ - that are very often nicely filled with extended radio emission \citep[see][and references therein]{BirzanCavities}. 
%Theoretical Challenges in Modeling AGN feedback 

So far, theoretical models of AGN feedback have focused on simulating purely hydrodynamic, kinematically supported jets. However, these simulations have so far failed to reproduce the observed X-ray morphology as depressions with enhanced rim emission and the thermodynamic structure of the bubbles, which often appear cooler in projection. 
In fact, bubbles in hydro simulations are very quickly Òshredded apartÓ by Rayleigh-Taylor instabilities on the top and Kelvin-Helmholtz instabilities on the sides. Several ideas have been put forward to prevent this from happening, with two mechanisms showing the most promise: viscosity and magnetic fields. Our work focuses on the second idea, in particular the extreme case of magnetically dominated bubbles.

\section{Mock {\it Chandra} Observations from Magnetically Dominated Jet Simulations}   %%% Top level 

We expand our earlier work on the first MHD simulations of magnetically dominated jets \citep{LiMHD1,LiMHD2, LiMHD3}. Our model only relies on the injection of non-force-free magnetic fields in the center of the cluster. These fields then expand and launch a Òmagnetic towerÓ. 
While doing so, the fields first self-collimate to form a current-carrying jet structure. When the ambient gas pressure starts dropping outside the X-ray core radius, the jet column then  expands, forming the lobe. During the subsonic lobe inflation process, the lobe pushes the hot gas outward, and evacuates a cavity into the cluster atmosphere. 

\begin{figure}
\begin{center}
\includegraphics[width=0.495\textwidth]{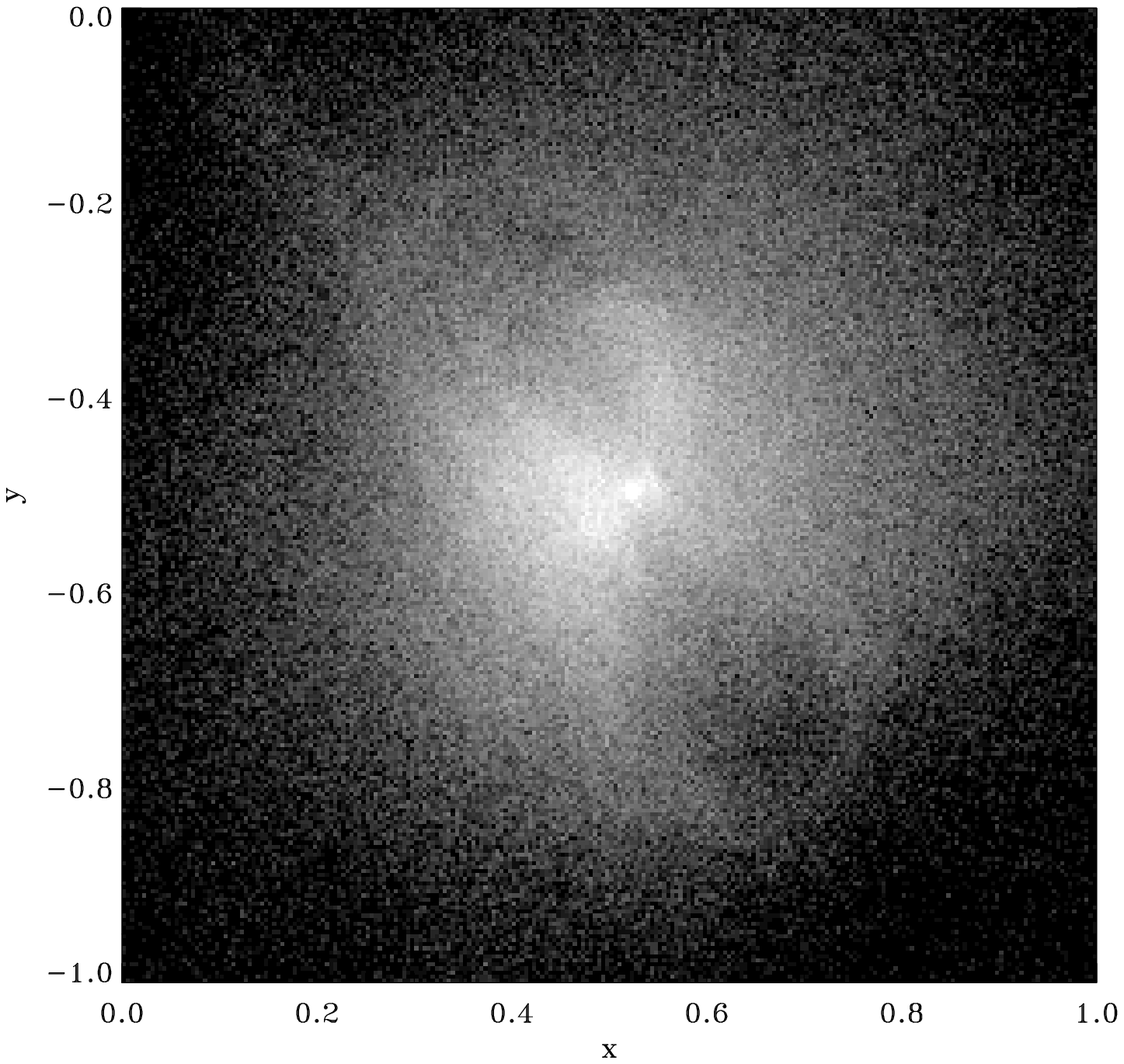}
\includegraphics[width=0.495\textwidth]{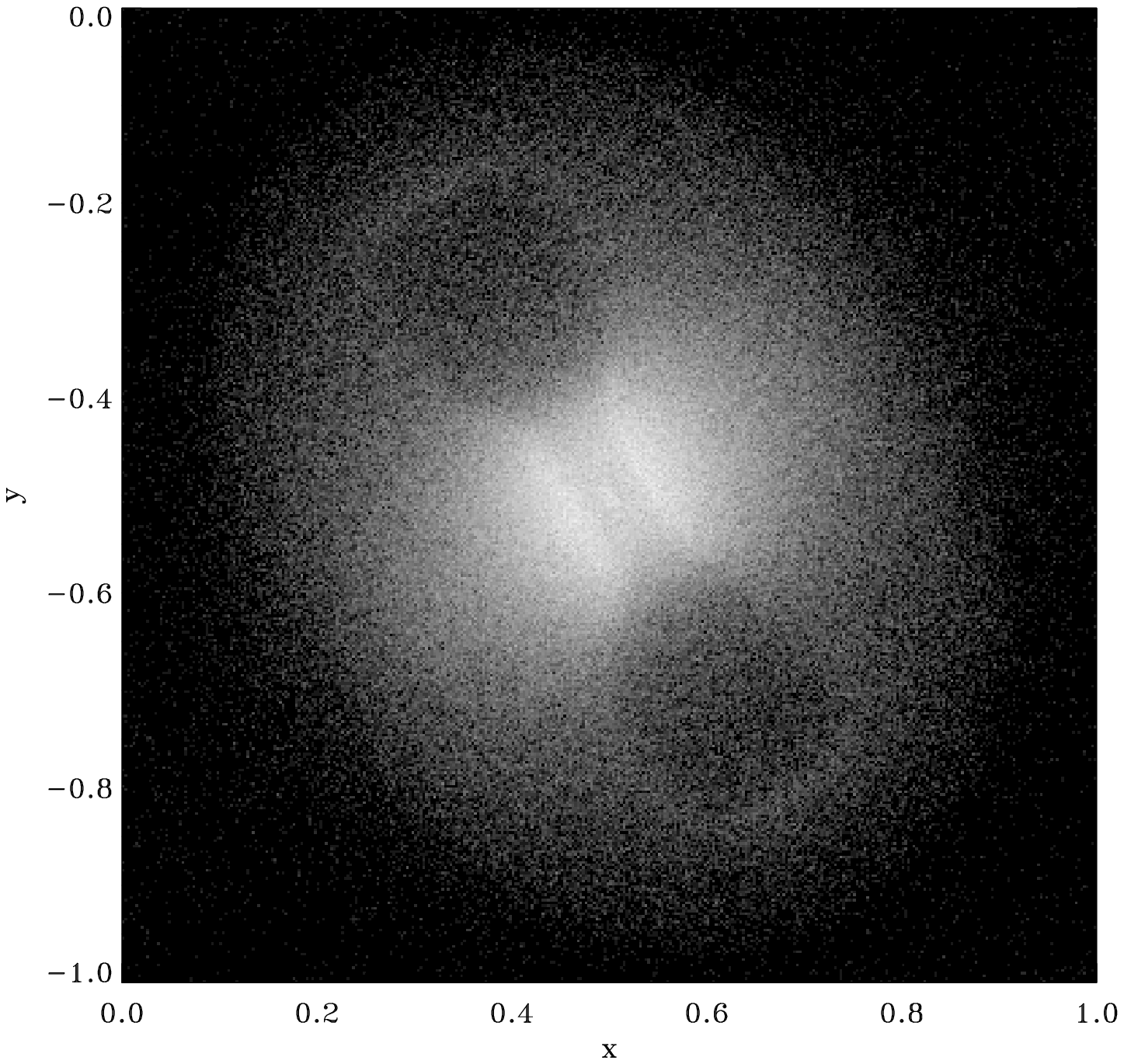}
\end{center}
\caption{Qualitative morphological comparison between an actual {\it Chandra} observation and a mock {\it Chandra} image from our simulation. Left: X-ray image of the core of {\it Hydra~A}; Right: Mock image of a simulated current-dominated MHD jet. Note the morphological similarities of depressed X-ray surface brightness at the location of the jet lobe, surrounded by enhanced rim emission. We emphasize that the rather wide width of the jet structure in the center is solely due to limited numerical resolution in this simulation. 
\label{f.mockchandra}}
\end{figure}

%Comparison of X-ray Features with Observation
We then produce mock Chandra X-ray observations with the simulation tool MARX. Figure \ref{f.mockchandra}b shows X-ray snapshots of the evolution of one simulations with 200,000 counts. 
Note the depressed X-ray emission at the the location of the lobe, surrounded by enhanced rim emission. The magnetic fields successfully stabilize the bubbles against disruption. These key morphological features are very close to actual cluster observations, shown for Hydra A in Figure \ref{f.mockchandra}a. 
In projection the slow, subsonic expansion of the bubbles letÕs them actually appear cooler than their surroundings. To our knowledge, no purely hydrodynamic jet simulation has yet been able to reproduce this observational fact. 

% DON'T FORGET TO CITE YOURSELF, IN ORDER TO AVOID SELF-PLAGIARISM!!!!!

\section{X-ray Cavity Sizes: Confronting Theory with Observations}   %%% Top level 

\subsection{Model Predictions: Adiabatic vs. Current Dominated}

\begin{figure}
\begin{center}
%\hspace{0.495\textwidth}
\includegraphics[width=0.495\textwidth]{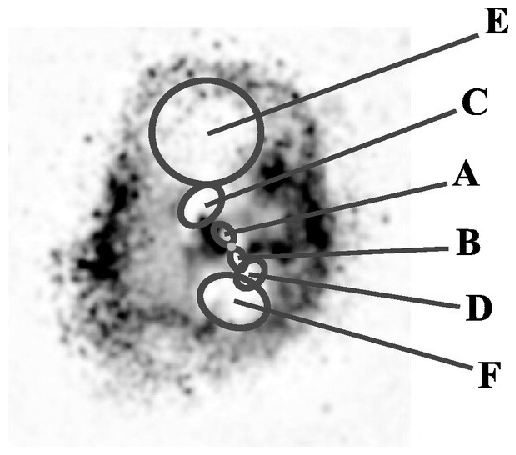}
\includegraphics[width=0.495\textwidth]{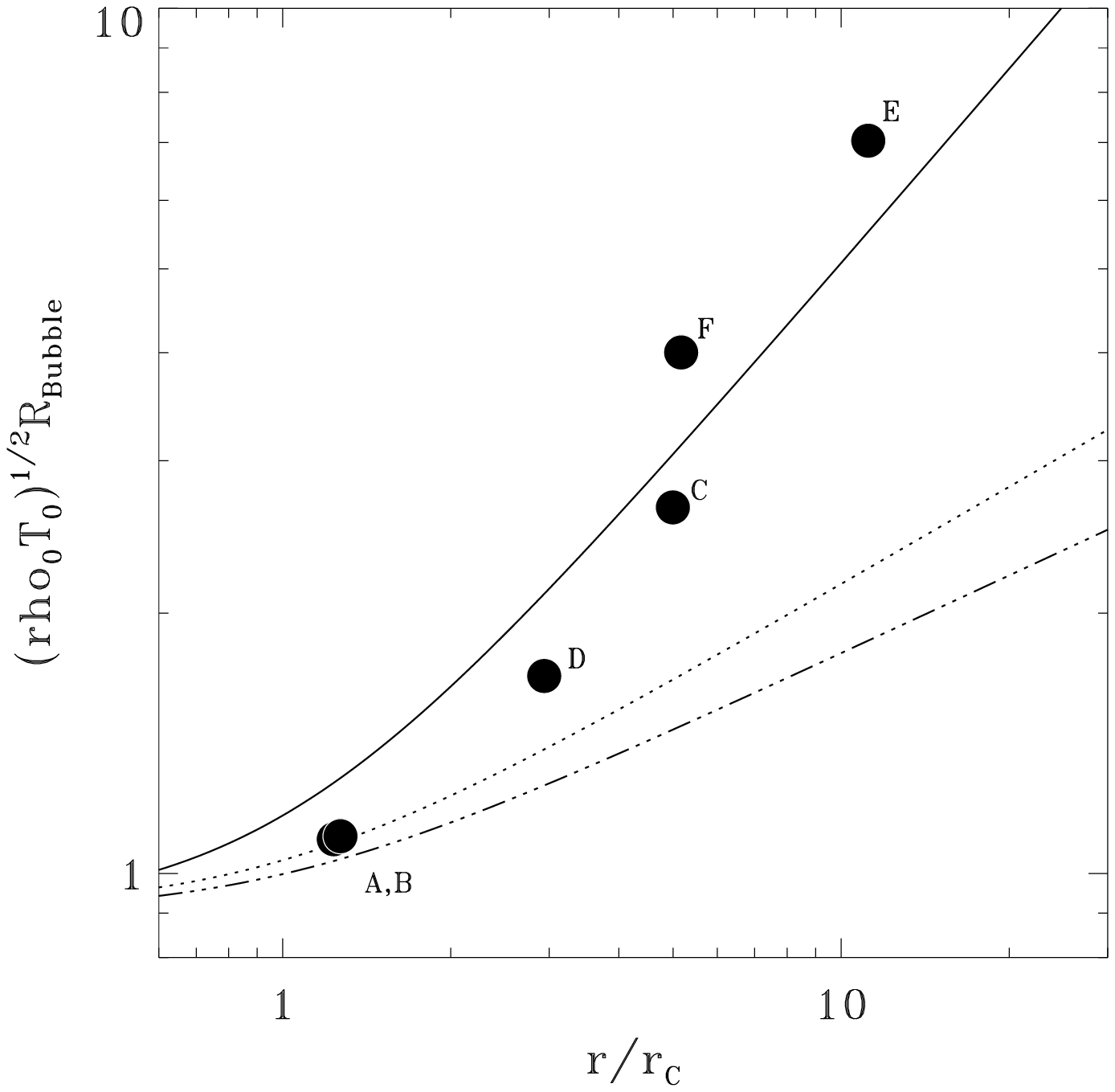}
%\vspace{5cm}
\end{center}
\caption{Left: Black colors show X-ray excess in {\it Chandra}'s {\it Hydra~A} image, when an elliptical model is subtracted; circles demonstrate the location of cavities A-F \citep[Figure reproduced from][with permission from the authors] {WiseHydraA}; Right: Bubble size $R_b$ as a function of bubble location $r$ for the cavities A-F. The lines show several model predictions. Triple-dot-dashed line: purely adiabatic expansion for a $\Gamma = 5/3$; dotted line: same, but for $\Gamma = 4/3$; solid line: magnetically dominated, current-carrying jet (our model). Note that the evolution of the bubble sizes can be explained by a single current running through the system, whereas the adiabatic expansion model would require subsequently stronger outbursts. This figure has been reproduced from \citet{DiehlMagBubbles}. \label{f.hydraa}}
\end{figure}
% DON'T FORGET TO CITE YOURSELF, IN ORDER TO AVOID SELF-PLAGIARISM!!!!!

Purely hydrodynamic bubble models predict a very simple behavior as they rise adiabatically in the intracluster medium (assumed to follow an isothermal $\beta$-model). Thus, -- assuming the bubble stays intact -- keeping $pV^\Gamma={\rm constant}$ gives a prediction of the bubble size as a function of radius $r$:
\begin{equation}
%  R_{b,\Gamma}=\left({ {3}\over{4\pi } }\right)^{1\over3} \left( {{\rho_0\, kT_0}\over{\eta\,\bar{m}} } \right)^{-{1\over3\Gamma} }         \left[ 1+ (r/r_c)^2 \right]^{{\beta\over2\Gamma}} 
  R_{b,\Gamma}=R_{b,0} \, \left[ 1+ (r/r_c)^2 \right]^{{\beta\over2\Gamma}},
\end{equation}
where $R_{b,0}$ denotes the fiducial bubble size projected back to the cluster center. 

For our current-carrying MHD simulations we get a very different prediction. As the lobes are magnetically dominated, the size of the cavity is determined by the point where the external gas pressure balances the internal magnetic pressure ($\propto B^2$). This B-field is mainly generated by the current $I_z$, and falls off as $I_z/R$ (AmpreÕs law), thus predicting:  
\begin{equation}
%  R_{b,{\rm I_z}}= I_z\, \left( {{8\pi\,\rho_0\, kT_0} }\over{\bar{m}} \right)^{-{1 \over 2}}         \left[ 1+ (r/r_c)^2 \right]^{{3\beta\over 4}} 
  R_{b,{\rm I_z}}= R_{b,0} \left[ 1+ (r/r_c)^2 \right]^{{3\beta\over 4}} 
\end{equation}
In this magnetically dominated case, the bubble size $R_{b,0}$ projected back to the center is proportional to the current $I_z$ that is being driven through the system. 

\subsection{A Perfect Test Case: The Multi-Cavity System {\it Hydra~A}}

Unfortunately, we cannot observe the same cavity multiple times as it rises in the cluster. However, cavities in the X-ray emission of clusters provide a unique fossil long-term record of past AGN activity. The multiple cavity system of Hydra A is a great example (Figure \ref{f.hydraa}a), consisting of 3 pairs of bubbles at various radii.
Assuming that each cavity pair was created by a similar outburst, we can then apply our formalism and test the predictions of the models. Figure \ref{f.hydraa}b shows the radial evolution of the appropriately scaled bubble sizes (circles). The red lines show the predictions for the adiabatic bubbles (for $\Gamma=4/3$ and $\Gamma=5/3$), which severely underestimates the bubble sizes at large radii. However, the prediction from the magnetically dominated bubbles, denoted by the green line, fits the data very well. In order to explain the large bubble sizes at large radii with purely hydrodynamic models, one has to invoke successively larger outbursts or continuous inflation.

\begin{figure}
\begin{center}
\includegraphics[width=0.43\textwidth]{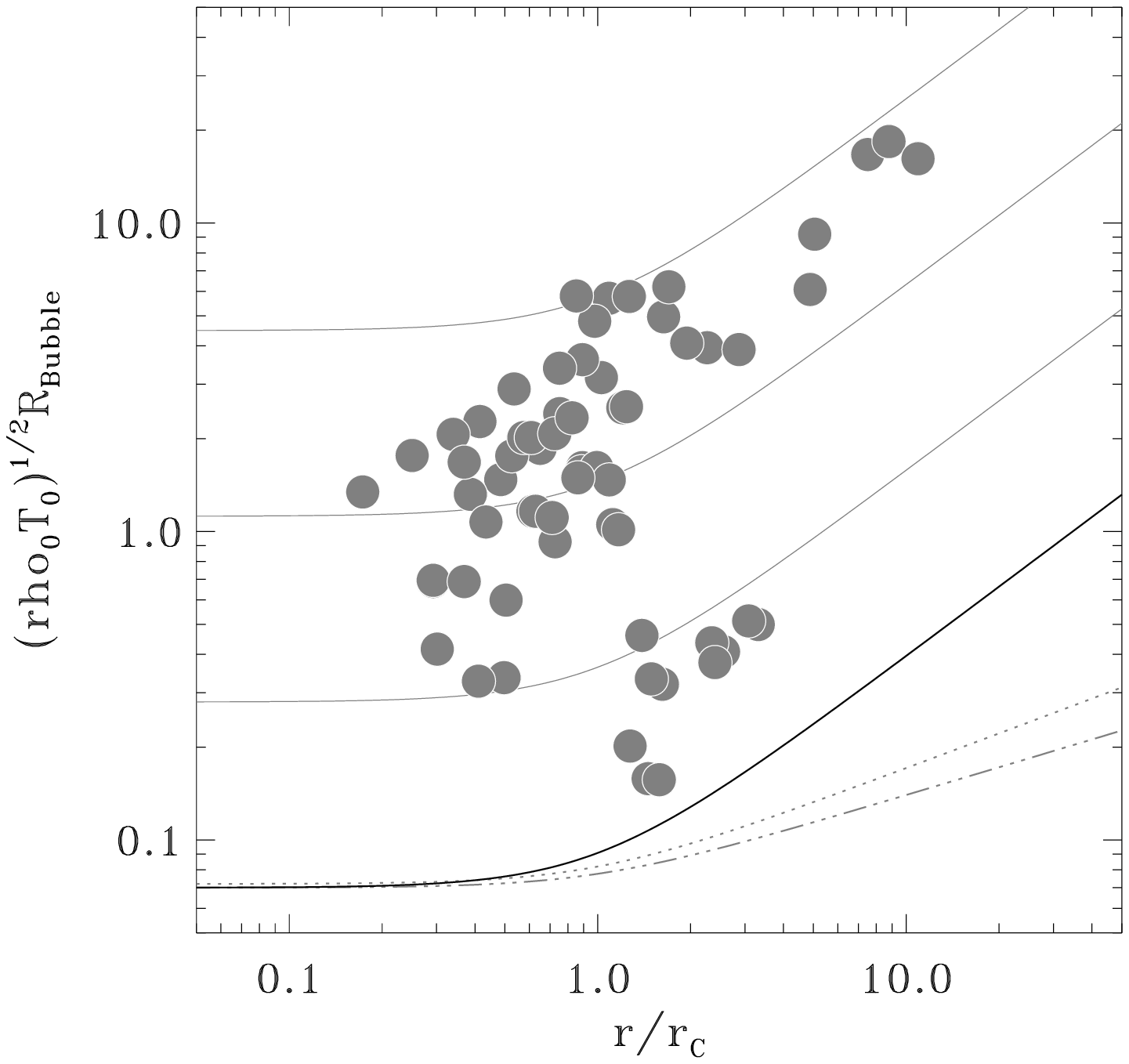}
\includegraphics[width=0.43\textwidth]{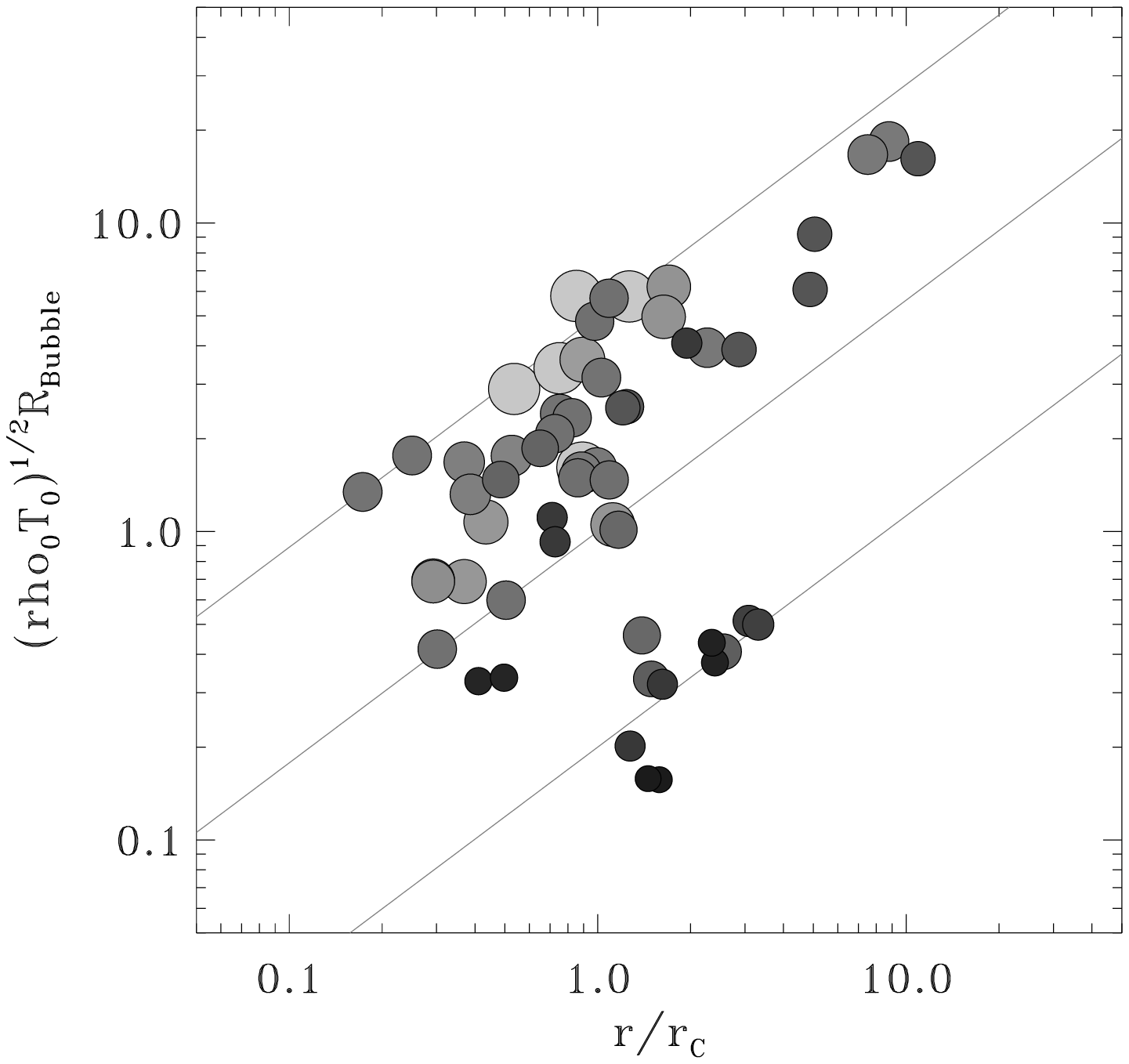}
\includegraphics[width=0.12\textwidth]{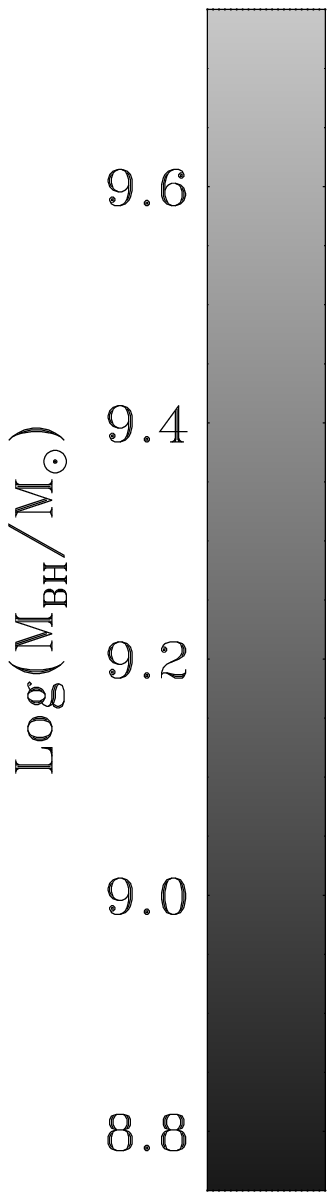}
%\vspace{5cm}
\end{center}
\caption{Left: Scaled bubble size $R_b$ as a function of bubble location $r$ for a sample of 64 cavities. The lines are the same as in Figure \ref{f.hydraa}b; Right: Same plot, but now symbols are colored according to their black hole masses. Larger symbols with lighter color indicate larger black hole masses. This figure has been reproduced from \citet{DiehlMagBubbles}.\label{f.bubblesize}}
\end{figure}

\subsection{Statistical Approach: A Large Cavity Sample}

Figure \ref{f.bubblesize} shows the same plot of scaled bubble size vs. radius for a much larger sample of 64 cavities. Note that we observe the same radial trend as for Hydra A. Cavities at larger radii are overly large compared to predictions from purely adiabatic bubbles. However, the trend nicely follows the general slope expected for our magnetically dominated model. 
It is also worth noting that this result is independent of adding viscosity to the adiabatic model, as viscosity cannot magically inflate the bubble. The grey lines denote lines of constant current, spaced by factors of 4, indicating that the width of the correlation can be reproduced by a narrow range in currents (about a factor of 30). Figure \ref{f.bubblesize}b shows the same plot but with symbol sizes indicating the central BH mass. We find that more massive BHs drive larger currents and inflate bigger cavities. 

While this result is suggestive, one has to keep one caveat in mind, namely incompleteness. \citet{EnsslinBubbledetectability} have tackled this problem by analytically computing the signal-to-noise ratio of a spherical void in a single-temperature $\beta$-model atmosphere. As we do not have an automatic bubble detection tool, but rather identify cavities manually by eye, it is difficult to estimate how incompleteness really affects our sample. Nevertheless, we can definitely say that magnetically dominated cavities will be much easier to identify, as they stay intact and expand much faster in general. To put a quantitative constraint on the nature of bubbles in cluster atmosphere, we need Monte-Carlo simulations of bubbles including incompleteness effects, which are currently underway and will be presented in a forthcoming paper \citep{DiehlMagBubbles}.

% DON'T FORGET TO CITE YOURSELF, IN ORDER TO AVOID SELF-PLAGIARISM!!!!!

\section{Conclusions}

We present results from magnetically dominated, current-carrying jet structures in clusters, and show that these jets subsonically inflate bubbles into the intracluster medium. We find that the helical magnetic field lines support these bubbles and stabilize them against Rayleigh-Taylor and Kelvin-Helmholtz instabilities. 

Mock Chandra observations show that the inflated bubbles morphologically and thermodynamically strongly resemble the cavities found in X-ray observations of clusters. In particular, we find enhanced rim emission surrounding the cavities, as well as an effective decrease in temperature at the bubble location. 

An analysis of bubble sizes in the multiple cavity system Hydra A, as well as in a large sample of 64 cavities in 31 clusters favors magnetically dominated cavities over purely hydrodynamic bubbles. An analysis of incompleteness effects will be addressed in an upcoming paper \citep{DiehlMagBubbles}.

Based on the assumption that our model is correct, we find that the current flowing in these systems depends on the central BH mass, offering a potential way to constrain BH mass from imaging X-ray observations.

%section head (remove "%" symbol)
%\subsection{}   %%% Second level section head (remove "%" symbol)
%\subsubsection{}   %%% Lowest level section head (remove "%" symbol)
%\section*{}    %%% Unnumbered top level section head (remove "%" symbol)
%\subsection*{}   %%% Unnumbered second level section head (remove "%" symbol)

%\acknowledgements %%% Text of acknowledgements runs on after this command.

%NOTE TO MYSELF: There is a problem with bibtex in this template. You have to edit the bbl file at the end and remove the "{3}" so it reads only \begin{thebibliography}{}
\bibliographystyle{apj}
\bibliography{../bibtex/allreferences.bib}

%%% THE BIBLIOGRAPHY
%%%
%%% CONSULT SECTION 3 OF "INSTRUCTIONS FOR AUTHORS" FOR HOW TO USE NATBIB.
%%% AUTHORS ARE ENCOURAGED TO USE EITHER THE "THEBIBLIOGRAPY" ENVIRONMENT
%%% BY UNCOMMENTING (DELETING THE "%" SYMBOL) THE COMMANDS BELOW, OR BY
%%% USING THE BIBTEX ENVIRONMENT. TO FIND OUT WHICH IS APPLICABLE TO YOUR
%%% CONTRIBUTION, CONSULT THE VOLUME EDITORS FOR YOUR PROCEEDINGS.
%%%

%\begin{thebibliography}{}
%\bibitem[LiMHD1]{testme}
%\bibitem[]{}
%\bibitem[]{}
%\bibitem[]{}
%\bibitem[]{}
%\bibitem[]{}
%\bibitem[]{}
%\bibitem[]{}
%\bibitem[]{}
%\bibitem[]{}
%\bibitem[]{}
%\bibitem[]{}
%\end{thebibliography}

\end{document}